\documentclass[pra,twocolumn,a4paper,groupedaddress,showpacs,floatfix,aps,10pt,nobibnotes]{revtex4-1}
\usepackage{graphicx,amsmath,amssymb,amsfonts,dsfont,subfigure,color}

\begin{document}

\title{Directional THz generation in hot Rb vapor excited to a Rydberg state}

\author{Mark Lam${}^{1}$}
\author{Sambit B. Pal${}^{1}$}
\author{Thibault Vogt${}^{1,2}$}
\author{Martin Kiffner${}^{1,3}$}
\author{Wenhui Li${}^{1,4}$}

\affiliation{Centre for Quantum Technologies, National University of Singapore, 3 Science Drive 2, Singapore 117543${}^1$}
\affiliation{MajuLab, CNRS-UNS-NUS-NTU International Joint Research Unit UMI 3654, Singapore 117543${}^2$}
\affiliation{Clarendon Laboratory, University of Oxford, Parks Road, Oxford OX1 3PU, United Kingdom${}^3$}
\affiliation{Department of Physics, National University of Singapore, Singapore 117542${}^4$}
%\affil[*]{Corresponding author: wenhui.li@nus.edu.sg}

%% To be edited by editor
% \dates{Compiled \today}

%\ociscodes{ (020.1670), (020.5780), (300.6210).}

%% To be edited by editor
% \doi{\url{http://dx.doi.org/10.1364/XX.XX.XXXXXX}}

\begin{abstract}
We optically excite $^{85}$Rb atoms in a heated vapor cell to a low-lying Rydberg state 10D$_{5/2}$ and observe directional terahertz (THz) beams at 3.3 THz and 7.8 THz. These THz fields are generated by amplified spontaneous emission from the 10D$_{5/2}$ state to the 11P$_{3/2}$ and 8F$_{7/2}$ states, respectively.
In addition, we observe ultraviolet (UV) light produced by four-wave mixing of optical pump lasers and the 3.3 THz field. We characterize the generated THz power over the detuning and power of pump lasers, and identify experimental conditions favoring THz and UV generation, respectively.
Our scheme paves a new pathway towards generating high-power narrow-band THz radiation.
\end{abstract}

%\setboolean{displaycopyright}{true}

\maketitle

\smallskip

%Introduction

Terahertz (THz) radiation (100~GHz - 30~THz) is the spectral region between the microwave and infrared frequency bands. It has numerous potential applications in diverse fields including astronomical observation, material optimization, telecommunication and medical imaging~\cite{thzroadmap,mittleman2017}. However, compared to well-established applications for radio frequency and optical waves, THz radiation has not been utilized to its full potential due to a lack of powerful and compact sources as well as fast and sensitive detectors.
In response to this, there has been a continued effort in developing THz sources using a wide range of physical systems and techniques~\cite{Lewis_thzreview,zhong2017,fulop2020}. Gas-phase THz lasers, like optical lasers, are based on population inversion between discrete quantum levels in gaseous media, and have been explored since the early days~\cite{gebbie1964,chang1970}. In recent years, there has been renewed interest in THz molecular lasers, which achieved improved compactness and tunability by using mid-infrared quantum cascade lasers as pump~\cite{Pagies2016,chevalier2019}.

Lately, Rydberg atoms have emerged as a new physical platform for realizing THz technologies. The energy spacings between atomic Rydberg levels scale as $1/n^3$, and their corresponding transition dipole moments scale as $n^2$, where $n$ is a principal quantum number~\cite{gallagher:ryd}. The manifold of Rydberg states thus exhibits a plethora of strong dipole transitions in the microwave and THz regimes.
This feature has been utilized for the sensitive detection of microwave fields and their efficient conversion to optical fields~\cite{sedlacek2012microwave,holloway2017rfmetrology,meyer2020,han2018}. More recently, applications of Rydberg atoms for electric field sensing and imaging have been extended to the THz region~\cite{Wade2017}, and real-time terahertz imaging using Rydberg atomic vapor was reported in~\cite{downes2020}.

In this letter, we experimentally observe the generation of THz beams by optically exciting a hot Rubidium vapour to a Rydberg state. Directional THz radiation is created from population inversion between Rydberg states, followed by amplified spontaneous emission (ASE), i.e., amplification via stimulated emission in free space.
This work significantly extends previous experiments~\cite{lam2019,gai2021} where the presence of THz radiation was inferred from the observed UV light generated by four-wave mixing. In our current experiment, generated THz fields are detected directly, which is achieved by a custom-built vacuum cell with a silicon window that transmits THz radiation.
We characterize the generated THz power over the detuning and power of pump lasers, and identify experimental conditions favoring THz and UV generation, respectively. We find that the generated THz power is of the order of several $\mu$W for moderate pump laser powers achievable with compact diode lasers.

\begin{figure}[t]
	\centering
    \includegraphics[width= 1 \linewidth]{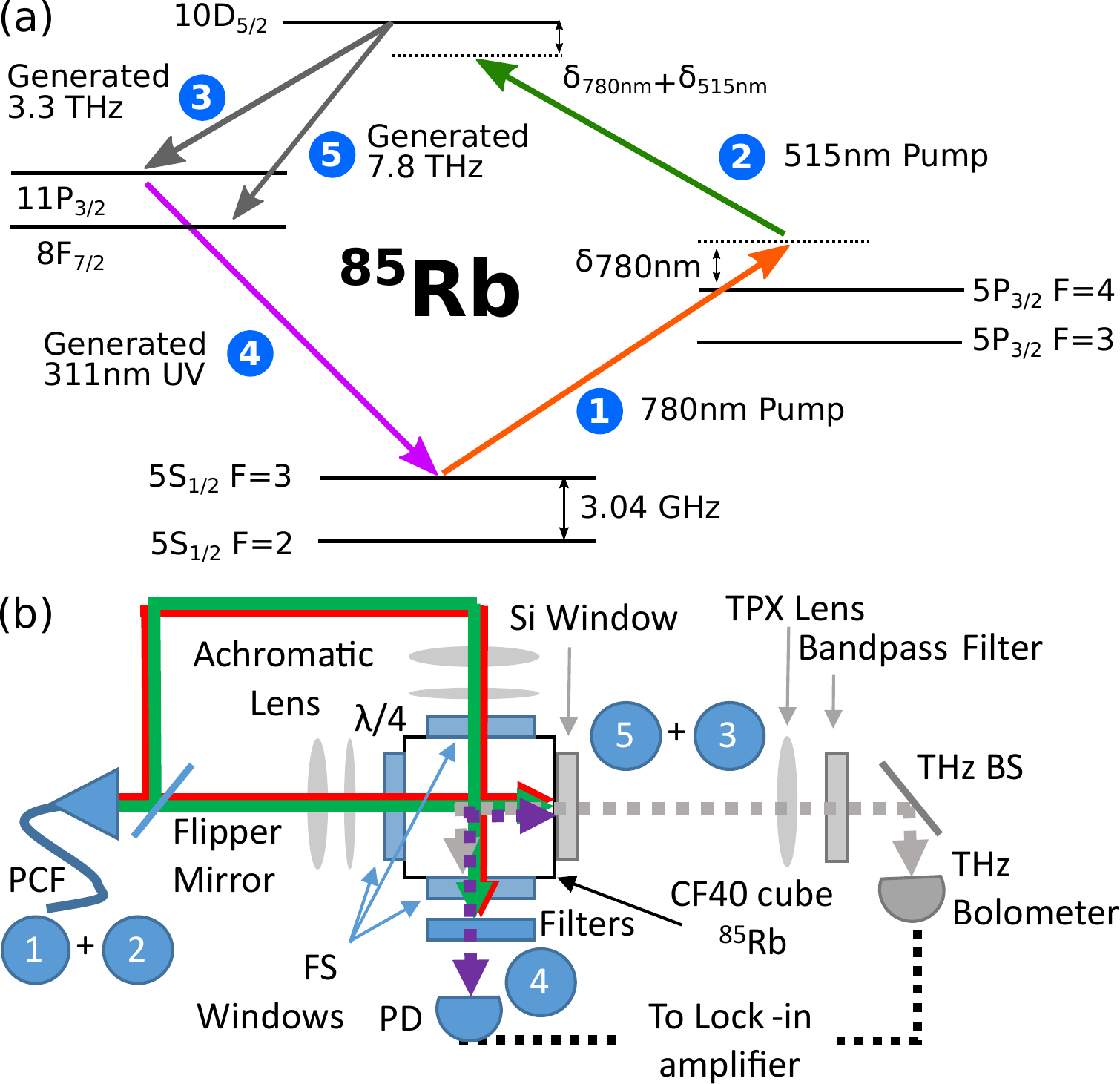}
	\caption{(a) Energy levels of $^{85}$Rb relevant to our experiment. $\delta_\textrm{780nm}$ and $\delta_\textrm{515nm}$ are the detunings of the two pump lasers from transitions 1 and 2, respectively. Numeric labels have been assigned to each transition for readers' convenience. (b) Schematics of the experimental setup. The two pump laser beams are denoted by the red and green lines, and selected optical elements are shown along beam paths. The vacuum vapor cell is sketched as a square with 4 labelled windows, while the THz beam is denoted by grey dotted line and the UV beam by purple dotted line. Details on beam paths and optical components along them are given in the text.}
	\label{scheme2}
\end{figure}

The relevant atomic energy levels in $^{85}$Rb  are shown in Fig.~\ref{scheme2}(a). The optical pump transitions 5S$_{1/2}$ F = 3 $\rightarrow$ 5P$_{3/2}$ F = 4 (transition 1) and 5P$_{3/2}$ F = 4 $\rightarrow$ 10D$_{5/2}$ (transition 2) are driven by two near-resonant laser fields at 780~nm and 515~nm, respectively.
The long radiative lifetime of the 10D$_{5/2}$ state ($\sim$780 ns)~\cite{beterov2009} enables population inversion with respect to the
lower-lying states with moderately strong pumping. The dipole-allowed transitions from the 10D$_{5/2}$ state to lower $n$P$_{3/2}$ and $n'$F$_{7/2}$ states, which have transition frequencies in the range of 1-100 THz, are listed in Table~\ref{dipolematrix}.
We experimentally measure the generated fields at 3.3 THz and 7.8 THz, from transitions 10D$_{5/2} \rightarrow$ 11P$_{3/2}$ (transition 3) and 10D$_{5/2} \rightarrow$ 8F$_{7/2}$ (transition 5), respectively. According to Table~\ref{dipolematrix}, these two transitions have by far the largest transition dipole-matrix elements, and should thus make the dominant contribution to the generated THz radiation.
As in our previous work~\cite{lam2019}, we also detect the UV light emitted on the 11P$_{3/2} \rightarrow$ 5S$_{1/2}$ transition (transition 4) via four-wave mixing.

%Experimental setup

Our experimental setup is illustrated in Fig.~\ref{scheme2}(b). We have built a vapor cell from a CF40 vacuum cube with an edge length of 7 cm, which is designed to measure the generated THz and UV radiation using two different optical paths as described below. Along the vertical axis of the cube (not shown in the figure), one port is connected to an atom source containing 1-gram pure rubidium sample with isotopic composition per its natural abundance; the other port is connected to a vacuum pump by which the chamber can be evacuated to around $10^{-8}$ mbar. Three of the four side ports are sealed with fused-silica (FS) windows, and the other one with a silicon (Si) window. The assembled cell is mounted in copper blocks connected to heatpipes wound with bifilar windings of heating wire, and maintained at a temperature of around 110 $^{\circ}$C, corresponding to a vapour density of around 1.1 $\times$ 10$^{13}$ cm$^{-3}$.

The two optical pump lasers (see~\cite{lam2019} for details) are overlapped using a photonic crystal fiber (PCF). The output beams of the PCF are switched with a flipper mirror between two possible paths for detecting either the generated UV or THz radiation. Along the two paths, the pump beams go through identical optical components and have the same beam configurations inside the vapor cell. Among different pump beam sizes, two configurations are reported here. In the ``collimated'' configuration, the pump beams are collimated over the length of the cell with the $1/e^2$ diameters $d_{515nm}$ = 1.1 mm for the 515~nm laser and $d_{780nm}$ = 2.2 mm for the 780~nm laser, respectively. In the ``focused'' configuration, both input beams are focused at the cell's geometric center with $d_{515nm}$ = 260 $\mu$m and $d_{780nm}$ = 700 $\mu$m. The pump beams are made circularly polarized by a quarter waveplate just before the cell.

\begin{table}[!t]
\centering
\begin{tabular}{c |c c c  |c |c c  c}
	\hline
	& \multicolumn{3}{c|}{10D$_{5/2} \rightarrow n$P$_{3/2}$}   &      & \multicolumn{3}{c} {10D$_{5/2} \rightarrow$ $n'$F$_{7/2}$}                             \\ \hline
	$n$  & $\nu$      & $\lambda$   & $|d_{\mathrm{DP}}|$   & $n'$   & $\nu$     & $\lambda$   & $|d_{\mathrm{DF}}|$     \\ \hline
	8 & 71.2             & 4.2         & 0.83                  & 5           & 88.4              &  3.4 & 0.65                         \\
	9 & 37.7            & 7.95         & 1.24                   & 6           & 48.0             &  6.25 & 1.65                         \\
	10  & 17.0            & 17.6         & 1.44                   & 7           & 23.6              &  12.7 & 4.82                            \\
	11  & 3.3            & 91.2         & 63.6                   & 8           & 7.8             &  38.7 & 27.68                           \\ \hline
	
\end{tabular}
\caption{Table of frequency $\nu$ (THz), wavelength $\lambda$ ($\mu$m), dipole matrix element $d_{DP}$ and $d_{DF}$ (ea$_0$) for transitions 10D$_{5/2} \rightarrow n$P$_{3/2}$ and 10D$_{5/2} \rightarrow$ $n'$F$_{7/2}$, where e is the elementary charge and a$_0$ is the Bohr radius. Further cascade transitions from $n$P and $n'$F states back to the 5S$_{1/2}$ state are not included here.} \label{dipolematrix}
\end{table}

\begin{table}[!b]
\centering
\begin{tabular}{c |c }
	\hline
	
	THz Optics     &  Efficiency at 3.3 (7.8) THz        \\ \hline
	
    Si Window      &  0.4 (0.4)                     \\
	TPX plate      & 0.55 (0.5)          \\
	BP filter (3.1-3.5 THz)         & 0.25 ($\sim$0)                     \\
    BP filter (6.7-8.1 THz)         & $\sim$0 (0.2)                     \\
	Beamsplitter   [reflection]       & 0.9 (0.9)                     \\ \hline
	
\end{tabular}
\caption{Transmission [reflection] efficiency of different THz components. All transmission percentages are experimentally obtained by measuring THz signal attenuation when an additional identical component is placed in the beam path. The bandpass filters' stop-band attenuations are verified by placing two dissimilar filters in the beam path. }   \label{THzTransmission}
\end{table}

The UV path goes through two FS windows that only transmit optical wavelengths. The generated UV is separated from the transmitted pump lights by UV-bandpass interference filters, and its power is measured with a calibrated UV-enhanced photodiode (PD).
In the THz path, the input port is a FS window as in the UV path. However, the output port is a Si window that is transparent for THz wave but opaque for the optical pump beams and generated UV light. A polymethylpentene (TPX) plano-convex lens is used to focus the THz radiation to be detected by a pyroelectric bolometer. The combination of the Si Window and the TPX lens allows the transmission of frequency up to 10 THz. In this way our setup allows us to measure the THz fields emitted on transitions 3 and 5 in Fig.~\ref{scheme2}(a). These two THz frequencies can be measured separately by using resonant bandpass filters made of copper mesh. An indium-tin-oxide-coated THz beamsplitter, which reflects THz wave but transmits infrared radiation, is used for steering the THz beam before the bolometer. The detailed characteristics of these THz components are given in Table~\ref{THzTransmission}. Note that the beam path after the TPX lens is enclosed and is additionally shielded from direct view of the heating elements to reduce any background effects, and an overall longer path is chosen to keep the detector further away from the heated vacuum chamber. Measurements of both UV and THz radiation are performed using the lock-in method, with modulation provided by a chopper wheel placed before the PCF.

%Experimental measurements - THz beam size

\begin{figure}[!t]
	\centering
	\includegraphics[width= 0.6 \linewidth]{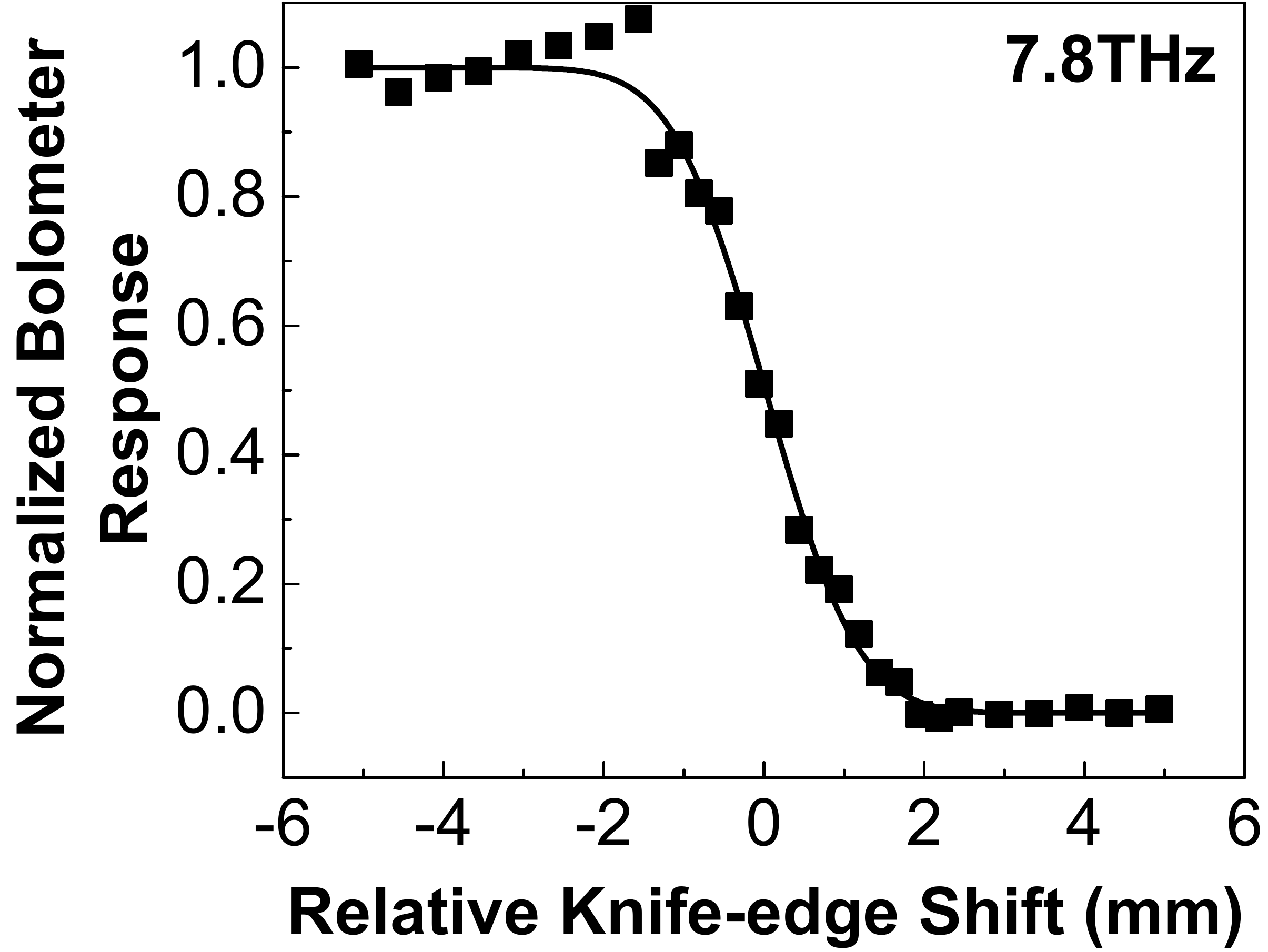}
	\caption{Knife-edge measurements of the generated THz beam at 3 cm from the exiting Si window.  The solid line is a fit to the data (black squares) with the error function, which is the integration of a Gaussian beam profile along the radial direction. The knife edge is placed after the Si window but before any other optics.}
	\label{knife}
\end{figure}

We verify the directionality of the generated THz radiation by measuring its transverse intensity distribution in a plane perpendicular to the propagation direction of the pump beams. Here the two input beams are under the ``collimated'' configuration described earlier. The result of a knife-edge measurement~\cite{Khosrofian:83} for the 7.8~THz beam is shown in Fig.~\ref{knife}. We find that the experimental data are fitted well by an error function, which demonstrates that the generated THz field has a localized Gaussian beam profile and propagates collinearly along the pump lasers.

From the measurement in Fig.~\ref{knife}, we estimate that the diameter of the 7.8~THz beam at the position of the Si window is approximately $d_{\text{7.8THz}}=1.5\pm 0.25\text{mm}$. This beam size is larger than the overlap area of the two pump beams, a cylinder with diameter $d_{515nm}$ = 1.1 mm. This result can be explained as follows. The THz radiation, generated by ASE, only requires excited atoms in the $10\text{D}_{5/2}$ state, but not necessarily the presence of both pump beams. That is, excited atoms can move outside the pump beams before ASE happens. Taking into account the long lifetime of the 10D$_{5/2}$ state, we calculate that the diameter of the active region where THz generation can take place is broadened to $\approx 1.4\text{mm}$ due to atomic motion. This is consistent with the above estimated THz beam size. We have also performed similar measurements on the 3.3~THz beam and find that it has a Gaussian profile as well. However, the lower power and larger divergence of the generated 3.3~THz radiation makes the measured sizes less accurate.

\begin{figure}[!t]
	\centering
		\includegraphics[width= 1 \linewidth]{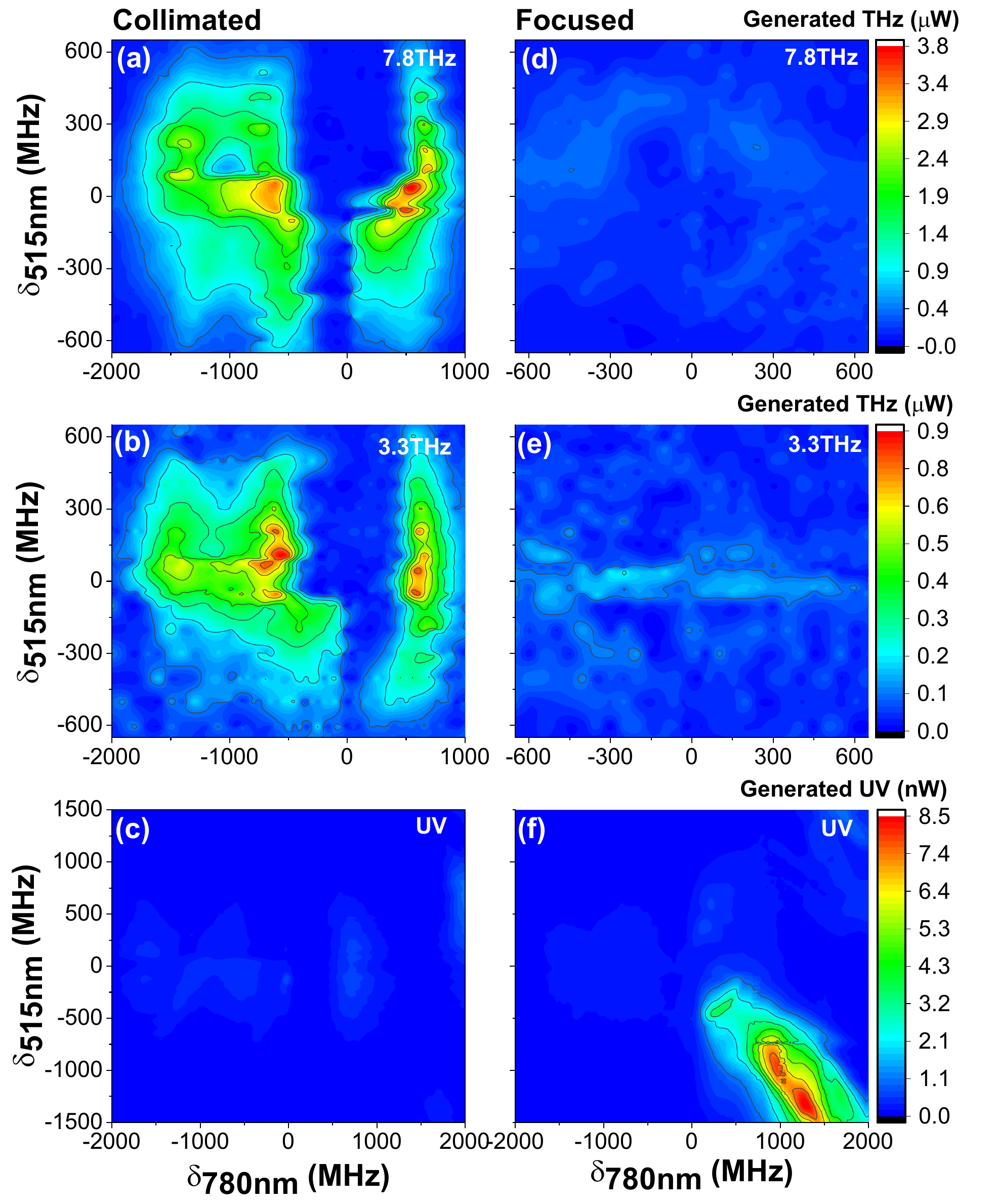}
\caption{Generated THz and UV powers plotted versus the two detunings $\delta_{780nm}$ and $\delta_{515nm}$ for two different beam configurations. The power of the 515 nm pump is $\sim$15 mW for both configurations; while the power of the 780 nm laser is $\sim$120 mW for the collimated configuration, but $\sim$160 mW for the focused configuration. The plotted THz powers are calibrated to be that just before the Si window inside the chamber by taking into account the efficiencies of various THz components listed in Table~\ref{THzTransmission}. The two spectra in each row share the same false-color bar right to them. }\label{thz-UVmultiplot}
\end{figure}

% Experimental measurements - the spectral features of THz generation.

We further characterize the spectral dependence of THz generation and present the spectra taken under the ``collimated'' and ``focused'' configurations in the left and right columns of Fig.~\ref{thz-UVmultiplot}, respectively. The different pump beam sizes in these two configurations lead to the differences in pump intensities and in active volumes where ASE can take place. More specifically, the pump beam intensities increase and the size of the active medium decreases when switching from the collimated to the focused configuration. This turns out to be the key factor that dictates the system to either favor THz or UV generation.

The THz spectra in Figs.~\ref{thz-UVmultiplot}(a) and~\ref{thz-UVmultiplot}(b) correspond to the collimated configuration and show vertical gaps around the resonance $\delta_{780nm}\sim0$. This is due to the strong absorption of the 780nm pump beam on the resonance of transition 1, which restricts excitation of atoms to the 10D$_{5/2}$ Rydberg state to a small volume near the entrance window. Moreover, we find that the maximal THz generation happens near $\delta_{515\text{nm}}\sim0$, and more THz power is generated in the red detuned region $\delta_{780nm} < 0$ than in $\delta_{780nm} > 0$. This latter observation is consistent with the fact that additional hyperfine resonances (e.g., 5S$_{1/2}$ F = 3 $\rightarrow$ 5P$_{3/2}$ F = 2 and 3) are located on the red-detuned side, which can enhance the excitation to the 10D$_{5/2}$ Rydberg level and consequently the THz generation

The THz spectra in Figs.~\ref{thz-UVmultiplot}(d) and~\ref{thz-UVmultiplot}(e) correspond to the focused configuration and show that the generated THz power is much smaller than in the collimated configuration [Figs.~\ref{thz-UVmultiplot}(a) and~\ref{thz-UVmultiplot}(b)]. Since measurements of the THz power vs. pump intensities show (see discussion of Fig.~\ref{thzpower} below) that ASE is already nearly saturated with the intensities in the collimated configuration, the increase of pump intensities in the focused beams will not enhance THz emission. Thus the decrease in generated THz power in the focused configuration can be largely attributed to the smaller area of the active region. The diameter of the active medium for ASE is about four times smaller for the focused configuration than that for the collimated configuration. The reduction of the peak 7.8~THz power in Fig.~\ref{thz-UVmultiplot}(d) compared to that in Fig.~\ref{thz-UVmultiplot}(a) is consistent with the corresponding active area difference, whereas the reduction for 3.3~THz in Fig.~\ref{thz-UVmultiplot} (e) is less pronounced.

The power of the generated UV light is shown in Figs.~\ref{thz-UVmultiplot}(c) and ~\ref{thz-UVmultiplot}(f) for the two pump beam configurations. Since the UV beam is generated via four-wave mixing (FWM)~\cite{lam2019}, it is expected to favor a more tightly focused beam configuration that provides the high intensities necessary for the nonlinear process. As phase-matching is required for the UV generation but not for THz generation, their spectral landscapes look different. The strong Kerr lensing at the red detuning side ($\delta_{780\text{nm}}<0$) distorts the 780~nm pump beam and destroys the phase-matching conditions~\cite{Vernier:10}. Therefore, the UV generation peak only appears at the blue-detuning side ($\delta_{780\text{nm}}>0$) and near two-photon resonance, as we had observed in our previous experiment~\cite{lam2019}.

\begin{figure}[t]
	\centering
	\includegraphics[width= 0.7 \linewidth]{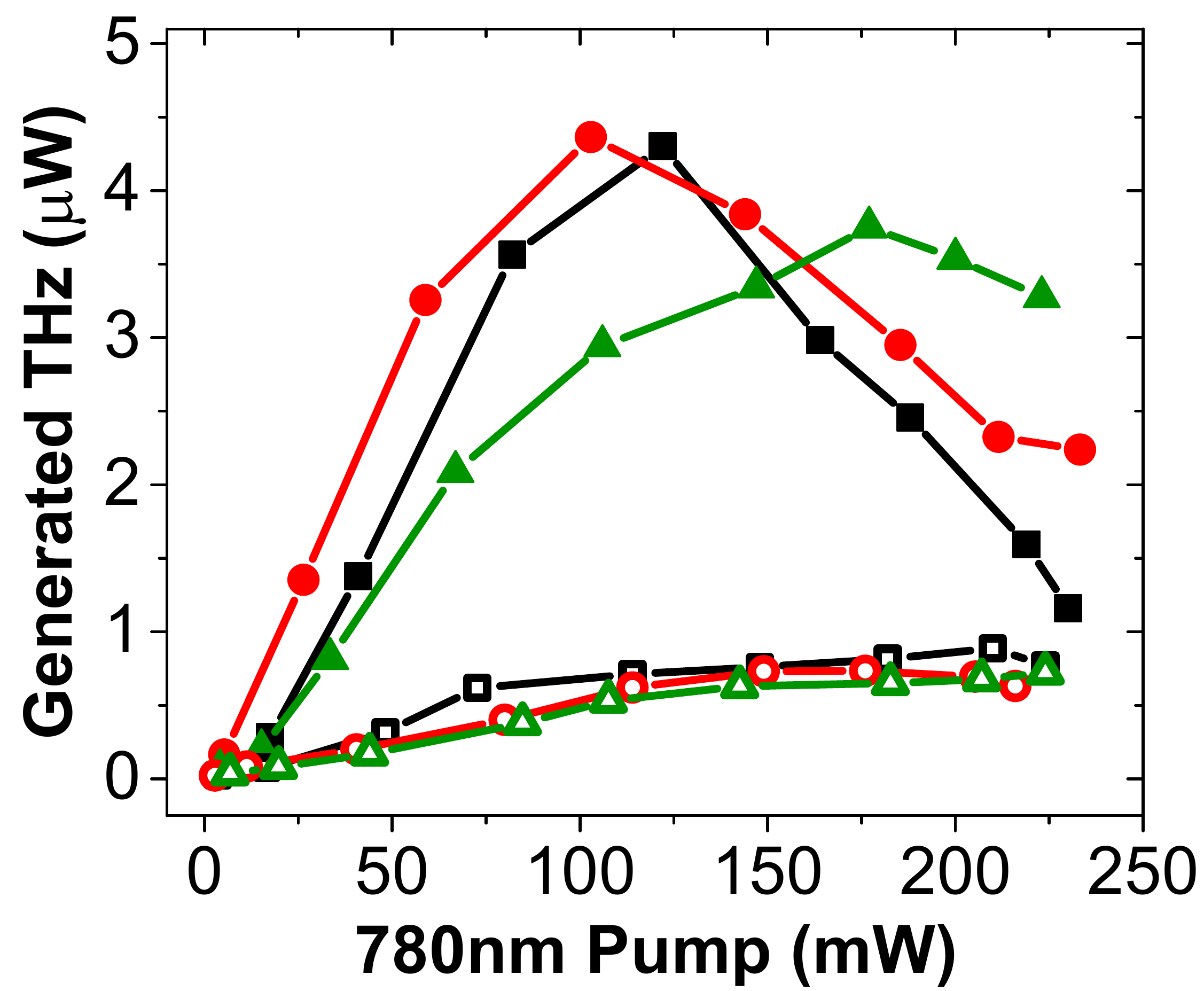}
	\caption{Generated powers of 3.3 THz (open symbol) and 7.8 THz (filled symbol) as a function of the 780 nm pump power for the 515 nm laser powers of 10 mW (green triangle), 12 mW (red circle), and 15 mW (black square). The input pump beams are in the collimated configuration. The detunings of both pump lasers are kept at the values that produce maximum THz power. The solid lines are a guide for the eye.}
	\label{thzpower}
\end{figure}

% Experimental measurements - pump power dependence.

Finally, we investigate the dependence of the generated THz radiation on the power of the pump lasers, and the result is shown in Fig.~\ref{thzpower}. The generated THz has only a weak dependence on the 515 nm power, and saturates when the power of the 515 nm beam is above 12 mW. The power of the 7.8~THz beam is always larger than that of the 3.3~THz beam. This observation is counterintuitive at first, since the dipole matrix element of the transition that emits the 3.3~THz radiation is more than twice as large as that for the 7.8~THz radiation (see Table~\ref{dipolematrix}). In order to investigate this further, we perform numerical simulations of Maxwell-Bloch equations~\cite{han2018} using a simple 4-level model. We find that for typical input parameters, the amplification of radiation does not grow exponentially but only linearly with the length of the medium. The reason for this is that the large dipole matrix elements in combination with the generated THz intensities result in very large Rabi frequencies (of the order of $100\text{MHz}$) that are well above the saturation threshold. In the saturated regime, the gain is independent of the incoming THz intensity and only depends on the number of excited atoms in the active region. If the two THz transitions were independent, our simple model predicts that the number of generated 3.3~THz and 7.8~THz photons is identical and thus the ratio of the generated  power is given by the frequency ratio $\omega_{7.8\text{THz}}/\omega_{3.3\text{THz}}\approx 2.4$. The measured power ratio is of that order of magnitude, and deviations are a consequence of complicated level structure and the fact that the two THz transitions influence each other since they both belong to the same atom. At high 780~nm power, the 3.3~THz radiation shows saturation; but the 7.8~THz power comes down. This decrease in THz generation is subject to further investigation.
%

%Conclusion
%\vspace{0.2cm}

In conclusion, we observe  narrow-band and directional THz beams with frequencies of 3.3 THz and 7.8 THz by optically pumping atoms into a low-lying Rydberg state. The power of the generated THz radiation is of the order of $\mu$W and can be further increased by increasing the diameter and power of the pump lasers, by enclosing the vapor cell in a cavity or by injecting the generated THz radiation into a quantum cascade laser (QCL) for amplification.
The observed THz radiation with high spectral brightness can be readily used for developing Rydberg-atom-based THz electrometry and THz imaging, which is currently limited to the sub-THz frequency range due to the lack of suitable sources.

\section*{Acknowledgments}

The authors acknowledge the support by the National Research Foundation, Prime Minister's Office, Singapore and the Ministry of Education, Singapore under the Research Centres of Excellence programme.

%\Bibliography



\begin{thebibliography}{21}%
\makeatletter
\providecommand \@ifxundefined [1]{%
 \@ifx{#1\undefined}
}%
\providecommand \@ifnum [1]{%
 \ifnum #1\expandafter \@firstoftwo
 \else \expandafter \@secondoftwo
 \fi
}%
\providecommand \@ifx [1]{%
 \ifx #1\expandafter \@firstoftwo
 \else \expandafter \@secondoftwo
 \fi
}%
\providecommand \natexlab [1]{#1}%
\providecommand \enquote  [1]{``#1''}%
\providecommand \bibnamefont  [1]{#1}%
\providecommand \bibfnamefont [1]{#1}%
\providecommand \citenamefont [1]{#1}%
\providecommand \href@noop [0]{\@secondoftwo}%
\providecommand \href [0]{\begingroup \@sanitize@url \@href}%
\providecommand \@href[1]{\@@startlink{#1}\@@href}%
\providecommand \@@href[1]{\endgroup#1\@@endlink}%
\providecommand \@sanitize@url [0]{\catcode `\\12\catcode `\$12\catcode
  `\&12\catcode `\#12\catcode `\^12\catcode `\_12\catcode `\%12\relax}%
\providecommand \@@startlink[1]{}%
\providecommand \@@endlink[0]{}%
\providecommand \url  [0]{\begingroup\@sanitize@url \@url }%
\providecommand \@url [1]{\endgroup\@href {#1}{\urlprefix }}%
\providecommand \urlprefix  [0]{URL }%
\providecommand \Eprint [0]{\href }%
\providecommand \doibase [0]{http://dx.doi.org/}%
\providecommand \selectlanguage [0]{\@gobble}%
\providecommand \bibinfo  [0]{\@secondoftwo}%
\providecommand \bibfield  [0]{\@secondoftwo}%
\providecommand \translation [1]{[#1]}%
\providecommand \BibitemOpen [0]{}%
\providecommand \bibitemStop [0]{}%
\providecommand \bibitemNoStop [0]{.\EOS\space}%
\providecommand \EOS [0]{\spacefactor3000\relax}%
\providecommand \BibitemShut  [1]{\csname bibitem#1\endcsname}%
\let\auto@bib@innerbib\@empty
%</preamble>
\bibitem [{\citenamefont {et~al.}(2017)}]{thzroadmap}%
  \BibitemOpen
  \bibfield  {author} {\bibinfo {author} {\bibfnamefont {S.~S.~D.}\
  \bibnamefont {et~al.}},\ }\href
  {http://stacks.iop.org/0022-3727/50/i=4/a=043001} {\bibfield  {journal}
  {\bibinfo  {journal} {Journal of Physics D: Applied Physics}\ }\textbf
  {\bibinfo {volume} {50}},\ \bibinfo {pages} {043001} (\bibinfo {year}
  {2017})}\BibitemShut {NoStop}%
\bibitem [{\citenamefont {Mittleman}(2017)}]{mittleman2017}%
  \BibitemOpen
  \bibfield  {author} {\bibinfo {author} {\bibfnamefont {D.~M.}\ \bibnamefont
  {Mittleman}},\ }\href@noop {} {\bibfield  {journal} {\bibinfo  {journal} {J.
  Appl. Phys.}\ }\textbf {\bibinfo {volume} {122}},\ \bibinfo {pages} {230901}
  (\bibinfo {year} {2017})}\BibitemShut {NoStop}%
\bibitem [{\citenamefont {Lewis}(2014)}]{Lewis_thzreview}%
  \BibitemOpen
  \bibfield  {author} {\bibinfo {author} {\bibfnamefont {R.~A.}\ \bibnamefont
  {Lewis}},\ }\href {http://stacks.iop.org/0022-3727/47/i=37/a=374001}
  {\bibfield  {journal} {\bibinfo  {journal} {Journal of Physics D: Applied
  Physics}\ }\textbf {\bibinfo {volume} {47}},\ \bibinfo {pages} {374001}
  (\bibinfo {year} {2014})}\BibitemShut {NoStop}%
\bibitem [{\citenamefont {Zhong}\ \emph {et~al.}(2017)\citenamefont {Zhong},
  \citenamefont {Shi}, \citenamefont {Xu}, \citenamefont {Liu}, \citenamefont
  {Wang}, \citenamefont {Mei}, \citenamefont {Yan}, \citenamefont {Fu},\ and\
  \citenamefont {Yao}}]{zhong2017}%
  \BibitemOpen
  \bibfield  {author} {\bibinfo {author} {\bibfnamefont {K.}~\bibnamefont
  {Zhong}}, \bibinfo {author} {\bibfnamefont {W.}~\bibnamefont {Shi}}, \bibinfo
  {author} {\bibfnamefont {D.}~\bibnamefont {Xu}}, \bibinfo {author}
  {\bibfnamefont {P.}~\bibnamefont {Liu}}, \bibinfo {author} {\bibfnamefont
  {Y.}~\bibnamefont {Wang}}, \bibinfo {author} {\bibfnamefont {J.}~\bibnamefont
  {Mei}}, \bibinfo {author} {\bibfnamefont {C.}~\bibnamefont {Yan}}, \bibinfo
  {author} {\bibfnamefont {S.}~\bibnamefont {Fu}}, \ and\ \bibinfo {author}
  {\bibfnamefont {J.}~\bibnamefont {Yao}},\ }\href@noop {} {\bibfield
  {journal} {\bibinfo  {journal} {Science China Technological Sciences}\
  }\textbf {\bibinfo {volume} {60}},\ \bibinfo {pages} {1801} (\bibinfo {year}
  {2017})}\BibitemShut {NoStop}%
\bibitem [{\citenamefont {F\"{u}l\"{o}p}\ \emph {et~al.}(2020)\citenamefont
  {F\"{u}l\"{o}p}, \citenamefont {Tzortzakis},\ and\ \citenamefont
  {Kampfrath}}]{fulop2020}%
  \BibitemOpen
  \bibfield  {author} {\bibinfo {author} {\bibfnamefont {J.~A.}\ \bibnamefont
  {F\"{u}l\"{o}p}}, \bibinfo {author} {\bibfnamefont {S.}~\bibnamefont
  {Tzortzakis}}, \ and\ \bibinfo {author} {\bibfnamefont {T.}~\bibnamefont
  {Kampfrath}},\ }\href@noop {} {\bibfield  {journal} {\bibinfo  {journal}
  {Advanced Optical Materials}\ }\textbf {\bibinfo {volume} {8}},\ \bibinfo
  {pages} {19900681} (\bibinfo {year} {2020})}\BibitemShut {NoStop}%
\bibitem [{\citenamefont {Gebbie}\ \emph {et~al.}(1964)\citenamefont {Gebbie},
  \citenamefont {Stone},\ and\ \citenamefont {Findlay}}]{gebbie1964}%
  \BibitemOpen
  \bibfield  {author} {\bibinfo {author} {\bibfnamefont {H.}~\bibnamefont
  {Gebbie}}, \bibinfo {author} {\bibfnamefont {N.}~\bibnamefont {Stone}}, \
  and\ \bibinfo {author} {\bibfnamefont {F.}~\bibnamefont {Findlay}},\
  }\href@noop {} {\bibfield  {journal} {\bibinfo  {journal} {Nature}\ }\textbf
  {\bibinfo {volume} {202}},\ \bibinfo {pages} {685} (\bibinfo {year}
  {1964})}\BibitemShut {NoStop}%
\bibitem [{\citenamefont {Chang}\ and\ \citenamefont
  {Bridges}(1970)}]{chang1970}%
  \BibitemOpen
  \bibfield  {author} {\bibinfo {author} {\bibfnamefont {T.}~\bibnamefont
  {Chang}}\ and\ \bibinfo {author} {\bibfnamefont {T.}~\bibnamefont
  {Bridges}},\ }in\ \href@noop {} {\emph {\bibinfo {booktitle} {Optics
  Communications}}},\ Vol.~\bibinfo {volume} {1}\ (\bibinfo  {publisher}
  {Elsevier},\ \bibinfo {year} {1970})\ pp.\ \bibinfo {pages}
  {423--426}\BibitemShut {NoStop}%
\bibitem [{\citenamefont {Pagies}\ \emph {et~al.}(2016)\citenamefont {Pagies},
  \citenamefont {Ducournau},\ and\ \citenamefont {Lampin}}]{Pagies2016}%
  \BibitemOpen
  \bibfield  {author} {\bibinfo {author} {\bibfnamefont {A.}~\bibnamefont
  {Pagies}}, \bibinfo {author} {\bibfnamefont {G.}~\bibnamefont {Ducournau}}, \
  and\ \bibinfo {author} {\bibfnamefont {J.-F.}\ \bibnamefont {Lampin}},\
  }\href@noop {} {\bibfield  {journal} {\bibinfo  {journal} {Appl. Phys.
  Lett.}\ }\textbf {\bibinfo {volume} {108}},\ \bibinfo {pages} {191106}
  (\bibinfo {year} {2016})}\BibitemShut {NoStop}%
\bibitem [{\citenamefont {Chevalier}\ \emph {et~al.}(2019)\citenamefont
  {Chevalier}, \citenamefont {Amirzhan}, \citenamefont {Wang}, \citenamefont
  {Piccardo}, \citenamefont {Johnson}, \citenamefont {Capasso},\ and\
  \citenamefont {Everitt}}]{chevalier2019}%
  \BibitemOpen
  \bibfield  {author} {\bibinfo {author} {\bibfnamefont {P.}~\bibnamefont
  {Chevalier}}, \bibinfo {author} {\bibfnamefont {A.}~\bibnamefont {Amirzhan}},
  \bibinfo {author} {\bibfnamefont {F.}~\bibnamefont {Wang}}, \bibinfo {author}
  {\bibfnamefont {M.}~\bibnamefont {Piccardo}}, \bibinfo {author}
  {\bibfnamefont {S.~G.}\ \bibnamefont {Johnson}}, \bibinfo {author}
  {\bibfnamefont {F.}~\bibnamefont {Capasso}}, \ and\ \bibinfo {author}
  {\bibfnamefont {H.~O.}\ \bibnamefont {Everitt}},\ }\href@noop {} {\bibfield
  {journal} {\bibinfo  {journal} {Science}\ }\textbf {\bibinfo {volume}
  {366}},\ \bibinfo {pages} {856} (\bibinfo {year} {2019})}\BibitemShut
  {NoStop}%
\bibitem [{\citenamefont {Gallagher}(1994)}]{gallagher:ryd}%
  \BibitemOpen
  \bibfield  {author} {\bibinfo {author} {\bibfnamefont {T.~F.}\ \bibnamefont
  {Gallagher}},\ }\href@noop {} {\emph {\bibinfo {title} {Rydberg Atoms}}}\
  (\bibinfo  {publisher} {Cambridge University Press},\ \bibinfo {address}
  {Cambridge},\ \bibinfo {year} {1994})\BibitemShut {NoStop}%
\bibitem [{\citenamefont {Sedlacek}\ \emph {et~al.}(2012)\citenamefont
  {Sedlacek}, \citenamefont {Schwettmann}, \citenamefont {K{\"u}bler},
  \citenamefont {L{\"o}w}, \citenamefont {Pfau},\ and\ \citenamefont
  {Shaffer}}]{sedlacek2012microwave}%
  \BibitemOpen
  \bibfield  {author} {\bibinfo {author} {\bibfnamefont {J.~A.}\ \bibnamefont
  {Sedlacek}}, \bibinfo {author} {\bibfnamefont {A.}~\bibnamefont
  {Schwettmann}}, \bibinfo {author} {\bibfnamefont {H.}~\bibnamefont
  {K{\"u}bler}}, \bibinfo {author} {\bibfnamefont {R.}~\bibnamefont {L{\"o}w}},
  \bibinfo {author} {\bibfnamefont {T.}~\bibnamefont {Pfau}}, \ and\ \bibinfo
  {author} {\bibfnamefont {J.~P.}\ \bibnamefont {Shaffer}},\ }\href@noop {}
  {\bibfield  {journal} {\bibinfo  {journal} {Nat. Phys.}\ }\textbf {\bibinfo
  {volume} {8}},\ \bibinfo {pages} {819} (\bibinfo {year} {2012})}\BibitemShut
  {NoStop}%
\bibitem [{\citenamefont {Holloway}\ \emph {et~al.}(2017)\citenamefont
  {Holloway}, \citenamefont {Simons}, \citenamefont {Gordon}, \citenamefont
  {Wilson}, \citenamefont {Cooke}, \citenamefont {Anderson},\ and\
  \citenamefont {Raithel}}]{holloway2017rfmetrology}%
  \BibitemOpen
  \bibfield  {author} {\bibinfo {author} {\bibfnamefont {C.~L.}\ \bibnamefont
  {Holloway}}, \bibinfo {author} {\bibfnamefont {M.~T.}\ \bibnamefont
  {Simons}}, \bibinfo {author} {\bibfnamefont {J.~A.}\ \bibnamefont {Gordon}},
  \bibinfo {author} {\bibfnamefont {P.~F.}\ \bibnamefont {Wilson}}, \bibinfo
  {author} {\bibfnamefont {C.~M.}\ \bibnamefont {Cooke}}, \bibinfo {author}
  {\bibfnamefont {D.~A.}\ \bibnamefont {Anderson}}, \ and\ \bibinfo {author}
  {\bibfnamefont {G.}~\bibnamefont {Raithel}},\ }\href@noop {} {\bibfield
  {journal} {\bibinfo  {journal} {IEEE Transactions on Electromagnetic
  Compatibility}\ }\textbf {\bibinfo {volume} {59}},\ \bibinfo {pages} {717}
  (\bibinfo {year} {2017})}\BibitemShut {NoStop}%
\bibitem [{\citenamefont {Meyer}\ \emph {et~al.}(2020)\citenamefont {Meyer},
  \citenamefont {Castillo}, \citenamefont {Cox},\ and\ \citenamefont
  {Kunz}}]{meyer2020}%
  \BibitemOpen
  \bibfield  {author} {\bibinfo {author} {\bibfnamefont {D.~H.}\ \bibnamefont
  {Meyer}}, \bibinfo {author} {\bibfnamefont {Z.~A.}\ \bibnamefont {Castillo}},
  \bibinfo {author} {\bibfnamefont {K.~C.}\ \bibnamefont {Cox}}, \ and\
  \bibinfo {author} {\bibfnamefont {P.~D.}\ \bibnamefont {Kunz}},\ }\href@noop
  {} {\bibfield  {journal} {\bibinfo  {journal} {J. Phys. B}\ }\textbf
  {\bibinfo {volume} {53}},\ \bibinfo {pages} {034001} (\bibinfo {year}
  {2020})}\BibitemShut {NoStop}%
\bibitem [{\citenamefont {Han}\ \emph {et~al.}(2018)\citenamefont {Han},
  \citenamefont {Vogt}, \citenamefont {Gross}, \citenamefont {Jaksch},
  \citenamefont {Kiffner},\ and\ \citenamefont {Li}}]{han2018}%
  \BibitemOpen
  \bibfield  {author} {\bibinfo {author} {\bibfnamefont {J.}~\bibnamefont
  {Han}}, \bibinfo {author} {\bibfnamefont {T.}~\bibnamefont {Vogt}}, \bibinfo
  {author} {\bibfnamefont {C.}~\bibnamefont {Gross}}, \bibinfo {author}
  {\bibfnamefont {D.}~\bibnamefont {Jaksch}}, \bibinfo {author} {\bibfnamefont
  {M.}~\bibnamefont {Kiffner}}, \ and\ \bibinfo {author} {\bibfnamefont
  {W.}~\bibnamefont {Li}},\ }\href@noop {} {\bibfield  {journal} {\bibinfo
  {journal} {Physical review letters}\ }\textbf {\bibinfo {volume} {120}},\
  \bibinfo {pages} {093201} (\bibinfo {year} {2018})}\BibitemShut {NoStop}%
\bibitem [{\citenamefont {Wade}\ \emph {et~al.}(2017)\citenamefont {Wade},
  \citenamefont {\u{S}ibali\'{c}}, \citenamefont {de~Melo}, \citenamefont
  {Kondo}, \citenamefont {Adams},\ and\ \citenamefont {Weatherill}}]{Wade2017}%
  \BibitemOpen
  \bibfield  {author} {\bibinfo {author} {\bibfnamefont {C.~G.}\ \bibnamefont
  {Wade}}, \bibinfo {author} {\bibfnamefont {N.}~\bibnamefont
  {\u{S}ibali\'{c}}}, \bibinfo {author} {\bibfnamefont {N.~R.}\ \bibnamefont
  {de~Melo}}, \bibinfo {author} {\bibfnamefont {J.~M.}\ \bibnamefont {Kondo}},
  \bibinfo {author} {\bibfnamefont {C.~S.}\ \bibnamefont {Adams}}, \ and\
  \bibinfo {author} {\bibfnamefont {K.~J.}\ \bibnamefont {Weatherill}},\ }\href
  {http://dx.doi.org/10.1038/nphoton.2016.214} {\bibfield  {journal} {\bibinfo
  {journal} {Nat. Photon.}\ }\textbf {\bibinfo {volume} {11}},\ \bibinfo
  {pages} {40} (\bibinfo {year} {2017})}\BibitemShut {NoStop}%
\bibitem [{\citenamefont {Downes}\ \emph {et~al.}(2020)\citenamefont {Downes},
  \citenamefont {MacKellar}, \citenamefont {Whiting}, \citenamefont
  {Bourgenot}, \citenamefont {Adams},\ and\ \citenamefont
  {Weatherill}}]{downes2020}%
  \BibitemOpen
  \bibfield  {author} {\bibinfo {author} {\bibfnamefont {L.~A.}\ \bibnamefont
  {Downes}}, \bibinfo {author} {\bibfnamefont {A.~R.}\ \bibnamefont
  {MacKellar}}, \bibinfo {author} {\bibfnamefont {D.~J.}\ \bibnamefont
  {Whiting}}, \bibinfo {author} {\bibfnamefont {C.}~\bibnamefont {Bourgenot}},
  \bibinfo {author} {\bibfnamefont {C.~S.}\ \bibnamefont {Adams}}, \ and\
  \bibinfo {author} {\bibfnamefont {K.~J.}\ \bibnamefont {Weatherill}},\
  }\href@noop {} {\bibfield  {journal} {\bibinfo  {journal} {Phys. Rev. X}\
  }\textbf {\bibinfo {volume} {10}},\ \bibinfo {pages} {011027} (\bibinfo
  {year} {2020})}\BibitemShut {NoStop}%
\bibitem [{\citenamefont {Lam}\ \emph {et~al.}(2019)\citenamefont {Lam},
  \citenamefont {Pal}, \citenamefont {Vogt}, \citenamefont {Gross},
  \citenamefont {Kiffner},\ and\ \citenamefont {Li}}]{lam2019}%
  \BibitemOpen
  \bibfield  {author} {\bibinfo {author} {\bibfnamefont {M.}~\bibnamefont
  {Lam}}, \bibinfo {author} {\bibfnamefont {S.~B.}\ \bibnamefont {Pal}},
  \bibinfo {author} {\bibfnamefont {T.}~\bibnamefont {Vogt}}, \bibinfo {author}
  {\bibfnamefont {C.}~\bibnamefont {Gross}}, \bibinfo {author} {\bibfnamefont
  {M.}~\bibnamefont {Kiffner}}, \ and\ \bibinfo {author} {\bibfnamefont
  {W.}~\bibnamefont {Li}},\ }\href@noop {} {\bibfield  {journal} {\bibinfo
  {journal} {Optics letters}\ }\textbf {\bibinfo {volume} {44}},\ \bibinfo
  {pages} {2931} (\bibinfo {year} {2019})}\BibitemShut {NoStop}%
\bibitem [{\citenamefont {Gai}\ \emph {et~al.}(2021)\citenamefont {Gai},
  \citenamefont {Liu}, \citenamefont {Wang}, \citenamefont {Chen},
  \citenamefont {Hu},\ and\ \citenamefont {Guo}}]{gai2021}%
  \BibitemOpen
  \bibfield  {author} {\bibinfo {author} {\bibfnamefont {B.}~\bibnamefont
  {Gai}}, \bibinfo {author} {\bibfnamefont {J.}~\bibnamefont {Liu}}, \bibinfo
  {author} {\bibfnamefont {P.}~\bibnamefont {Wang}}, \bibinfo {author}
  {\bibfnamefont {Y.}~\bibnamefont {Chen}}, \bibinfo {author} {\bibfnamefont
  {S.}~\bibnamefont {Hu}}, \ and\ \bibinfo {author} {\bibfnamefont
  {J.}~\bibnamefont {Guo}},\ }\href@noop {} {\bibfield  {journal} {\bibinfo
  {journal} {J. Quant. Spectrosc. Radiat. Transf}\ }\textbf {\bibinfo {volume}
  {258}},\ \bibinfo {pages} {107351} (\bibinfo {year} {2021})}\BibitemShut
  {NoStop}%
\bibitem [{\citenamefont {Beterov}\ \emph {et~al.}(2009)\citenamefont
  {Beterov}, \citenamefont {Ryabtsev}, \citenamefont {Tretyakov},\ and\
  \citenamefont {Entin}}]{beterov2009}%
  \BibitemOpen
  \bibfield  {author} {\bibinfo {author} {\bibfnamefont {I.}~\bibnamefont
  {Beterov}}, \bibinfo {author} {\bibfnamefont {I.}~\bibnamefont {Ryabtsev}},
  \bibinfo {author} {\bibfnamefont {D.}~\bibnamefont {Tretyakov}}, \ and\
  \bibinfo {author} {\bibfnamefont {V.}~\bibnamefont {Entin}},\ }\href@noop {}
  {\bibfield  {journal} {\bibinfo  {journal} {Physical review A}\ }\textbf
  {\bibinfo {volume} {79}},\ \bibinfo {pages} {052504} (\bibinfo {year}
  {2009})}\BibitemShut {NoStop}%
\bibitem [{\citenamefont {Khosrofian}\ and\ \citenamefont
  {Garetz}(1983)}]{Khosrofian:83}%
  \BibitemOpen
  \bibfield  {author} {\bibinfo {author} {\bibfnamefont {J.~M.}\ \bibnamefont
  {Khosrofian}}\ and\ \bibinfo {author} {\bibfnamefont {B.~A.}\ \bibnamefont
  {Garetz}},\ }\href {\doibase 10.1364/AO.22.003406} {\bibfield  {journal}
  {\bibinfo  {journal} {Appl. Opt.}\ }\textbf {\bibinfo {volume} {22}},\
  \bibinfo {pages} {3406} (\bibinfo {year} {1983})}\BibitemShut {NoStop}%
\bibitem [{\citenamefont {Vernier}\ \emph {et~al.}(2010)\citenamefont
  {Vernier}, \citenamefont {Franke-Arnold}, \citenamefont {Riis},\ and\
  \citenamefont {Arnold}}]{Vernier:10}%
  \BibitemOpen
  \bibfield  {author} {\bibinfo {author} {\bibfnamefont {A.}~\bibnamefont
  {Vernier}}, \bibinfo {author} {\bibfnamefont {S.}~\bibnamefont
  {Franke-Arnold}}, \bibinfo {author} {\bibfnamefont {E.}~\bibnamefont {Riis}},
  \ and\ \bibinfo {author} {\bibfnamefont {A.~S.}\ \bibnamefont {Arnold}},\
  }\href {\doibase 10.1364/OE.18.017020} {\bibfield  {journal} {\bibinfo
  {journal} {Opt. Express}\ }\textbf {\bibinfo {volume} {18}},\ \bibinfo
  {pages} {17020} (\bibinfo {year} {2010})}\BibitemShut {NoStop}%
\end{thebibliography}
\end{document}